\documentclass[11pt]{article}

\usepackage[english]{babel}

\usepackage[margin=1in]{geometry}
\usepackage{comment}
\usepackage{caption}
\usepackage{subcaption}
\usepackage{cite}
\usepackage{amsmath, amssymb}
\usepackage{graphicx}
\graphicspath{{./}{figures/}}
\usepackage{xcolor}
\usepackage[colorlinks=true, allcolors=blue]{hyperref}
\usepackage{subcaption}

\title{Effect of Compositional Fluctuation on\\
the Survival of Bet-hedging Species}
\author{Xiao Zhou \& BingKan Xue}
\date{\small Department of Physics, University of Florida, Gainesville, FL, United States}

\begin{document}

\maketitle

\begin{abstract}
Understanding the coexistence of diverse species in a changing environment is an important problem in community ecology. Bet-hedging is a strategy that helps species survive in such changing environments. However, studies of bet-hedging have often focused on the expected long-term growth rate of the species by itself, neglecting competition with other coexisting species. Here we study the extinction risk of a bet-hedging species in competition with others. We show that there are three contributions to the extinction risk. The first is the usual demographic fluctuation due to stochastic reproduction and selection processes in finite populations. The second, due to the fluctuation of population growth rate caused by environmental changes, may counterintuitively reduce the extinction risk for small populations. Besides those two, we reveal a third contribution, which is unique to bet-hedging species that diversify into multiple phenotypes: The phenotype composition of the population will fluctuate over time, resulting in increased extinction risk. We compare such compositional fluctuation to the demographic and environmental contributions, showing how they have different effects on the extinction risk depending on the population size, generation overlap, and environmental correlation.
\end{abstract}

\section{Introduction}

A central problem in ecology is to understand how species coexist in biological communities. In order to coexist, a species must be able to survive in the presence of other species despite potential competition for resources. Different mechanisms have been proposed to allow species coexistence, such as interspecific trade-offs in utilizing different resources \cite{Tilman2011, levins1979coexistence}, and environmental changes that prevent competitive exclusion by keeping ecosystems away from equilibrium \cite{hutchinson:1961}. More recent studies have analyzed the effect of fluctuating environments on species coexistence, such as the storage effect \cite{chesson1981environmental, chessen:2000, Chesson2018}, the time-averaged neutral theory of biodiversity \cite{Kalyuzhny2015, DANINO2016155, Hidalgo2017}, and population genetics in fluctuating environments \cite{Ashcroft2014, Cvijovic2015, melbinger2015impact}. Such studies are important as environmental fluctuation can pose additional challenges on a species's ability to survive in a community.

However, many of these studies modeled environmental fluctuation simply as causing the fitness of species to vary over time. Absent from these studies is that the species can develop strategies for adapting to such environmental variation. A common strategy used by many organisms to survive in varying environments is bet-hedging \cite{Seger1987, philippi1989hedging}. It is a risk-spreading strategy that counters environmental uncertainty by generating phenotypic variation in the population, even among individuals of the same genotype. Such phenotypic variation allows the population to have some individuals that are adapted to the environment at any given time, at the cost of having maladapted individuals at the same time. This helps reduce temporal variation of the population mean fitness, and thus alleviate the influence of unpredictable environmental changes. Many examples of bet-hedging involve a dormant physiological state of the organisms that helps them endure erratic periods of stressful environments, such as seed banks of annual plants \cite{cohen1966optimizing, Venable2007}, diapause of insect lavae \cite{Rajon2014}, and persister cells of bacteria \cite{Kussell2005}. By having dormant and active individuals both present in the population, the species can grow in good environmental conditions while being buffered from devastating down turns if a stressful condition hits.

How well the bet-hedging strategy helps a species survive depends on the statistics of environmental fluctuations and the composition of phenotypes in the population. Previous studies of bet-hedging often modeled a single species and focused on its average growth rate over time as the environment fluctuates. The optimal phenotype composition is considered to be the one that maximizes the expected long-term growth rate (or geometric mean fitness) of a population \cite{cohen1966optimizing, philippi1989hedging}. However, such calculations implicitly assume that the population continues to grow in the long term. In reality, the population size will be constrained by available resources and long-term growth cannot be sustained. What is important for the population is the ability to survive in competition with other species. Although there have been studies of bet-hedging that modeled multiple species (e.g., \cite{King2007, libby2019shortsighted}), few considered finite population sizes (but see \cite{Adler2008, Pande2020}).

In this paper, we will study how well bet-hedging species compete with others by finding the factors that affect the survival probability of a bet-hedging species. We will take into account both demographic and environmental fluctuations. For demographic fluctuation, there are classic results from population genetics that describe how the fixation probability depends on the relative fitness and population sizes \cite{kimura1962probability}. On the other hand, environmental fluctuation has been proposed to favor the relatively small population, thus promoting species coexistence \cite{chesson1981environmental}. In addition to these two contributions, there is a yet unstudied type of fluctuation that affects bet-hedging species: Even though the population tends to diversify according to an ideal phenotype distribution, the actual composition of the population may deviate from it because of finite population size. We will study how these different types of fluctuations affect the survival of bet-hedging species when competing with others. Besides recovering known results on demographic and environmental fluctuations, we show that compositional fluctuation generally undermines the bet-hedging species by increasing the extinction risk. This would induce a cost for the population to diversify in phenotypes, which offsets the commonly perceived evolutionary benefit of the bet-hedging strategy.

\section{Bet-hedging species in competition}

To study how a bet-hedging species compete with other species in a fluctuating environment, we consider the simplest situation that involves two species, one or both using a bet-hedging strategy. The total population size of the two species is assumed to be a constant, $N$. We will follow the general scheme of the Wright-Fisher model and consider discrete, non-overlapping generations, such that all individuals are replaced after every generation. New individuals in the next generation are sampled from the previous generation proportionally to their fitness values. The two species share the same environment that fluctuates over time, which affects the fitness of one or both species.

A bet-hedging species will diversify into multiple phenotypes that have different fitness in each environmental state. This is described by a fitness matrix $f_{ij}$, which specifies the fitness of a phenotype $\phi_{j}$ in an environmental state $\varepsilon_{i}$. We will focus on the case where there are two environmental states and two phenotypes. The environmental state in each generation is drawn independently with probability $p_i$. Upon birth, each individual takes one of the phenotypes according to a probability distribution $q_j$. When forming the next generation, individuals will be selected based on the fitness values of their own phenotypes in the current environment. Their offspring will have phenotypes drawn from the distribution $q_j$ independently of the parents. (Mathematical details of the model are described in the Appendix.) For the examples shown in this paper, we use a fitness matrix $f_{ij} = \left( \begin{smallmatrix} 1 & 0.2 \\ 1 & 5 \end{smallmatrix} \right)$. The fitness of one phenotype ($\phi_A$, first column) is the same in both environmental states, which represents a dormant state that is unaffected by environmental variation. The other phenotype ($\phi_B$, second column) has a low fitness in one environment ($\varepsilon_X$, top row) and high fitness in the other ($\varepsilon_Y$, bottom row), which represents a phenotype that produces a high yield under good conditions but is strongly affected by environmental stress. The two phenotypes are chosen to have the same geometric mean fitness when the environmental probabilities are equal, $p_i = (0.5, 0.5)$. The results we present in this paper are not sensitive to these parameter choices.

The probability distribution $q_j$ that characterizes the bet-hedging strategy will be parametrized by $(q, 1-q)$. We would like to evaluate the performance of such strategies in competition with another species. For example, we can simulate a bet-hedging species using a particular strategy $q$ to compete with a non-bet-hedging species with a constant fitness value. Each simulation starts with the same fraction of the bet-hedging species in the total population, and runs until one of the species goes extinct (i.e., the other species reaches fixation). After repeating the simulation many times, we estimate the probability that the bet-hedging species survives, and use this probability to measure the performance of the bet-hedging species. We have also developed numerical methods to calculate this fixation probability (see Appendix).

\begin{figure}
\centering
\includegraphics[width=0.5\textwidth]{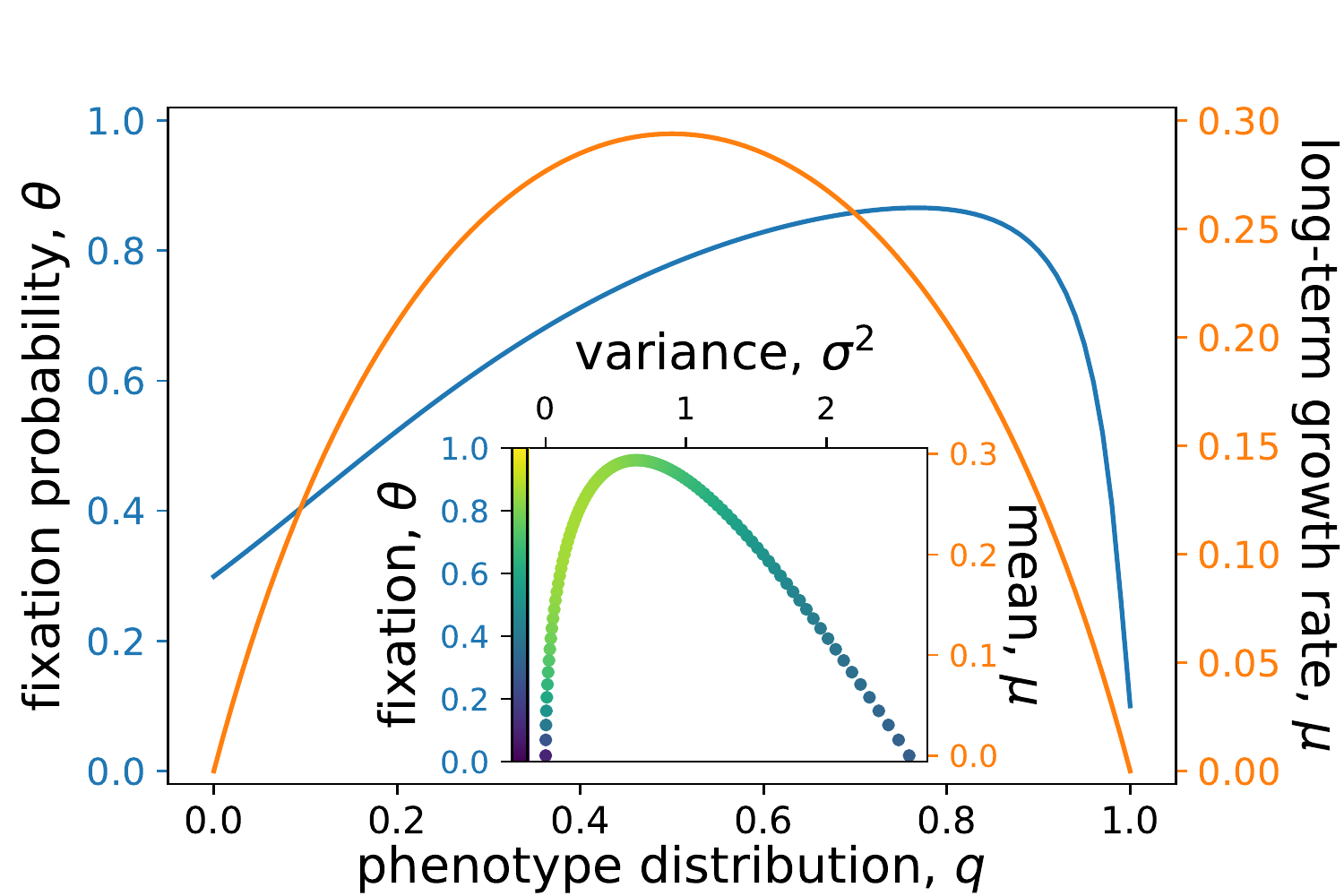}
\caption{\small Performance of a bet-hedging species competing with another species with constant fitness value $f_2 = 1$ (total population size $N = 100$, initial fraction of bet-hedging species $x = 0.1$). Main figure: fixation probability $\theta$ (blue) and long-term growth rate $\mu$ (orange) as a function of phenotype distribution $q$. Inset: mean ($\mu$) versus variance ($\sigma^2$) of the growth rate over time; the color of the points represents the fixation probability $\theta$. The $q$ that maximizes $\theta$ is different from that for maximum $\mu$ and has a smaller variance $\sigma^2$.}
\label{fig:EF}
\end{figure}

An example is shown in Figure~\ref{fig:EF}, where we plot the fixation probability $\theta$ of the bet-hedging species in a competition, as well as the expected long-term population growth rate $\mu$ of the species by itself, as a function of $q$. It can be seen that the location for the maximum fixation probability does not coincide with that for the largest long-term growth rate. Therefore, the long-term growth rate does not provide a good criterion for evaluating the persistence of the bet-hedging species. One reason is that the long-term growth rate does not account for fluctuations in the population growth rate over time. This can be visualized by a mean-variance plot (Figure~\ref{fig:EF} inset), where the y-axis shows the expected long-term growth rate given by:
\begin{equation} \label{eq:mu}
\mu = \sum_{i} p_i \log \Big( \sum_{j} f_{ij} \, q_j \Big) \equiv \sum_{i} p_i \log \bar{f}_i \,,
\end{equation}
and the x-axis shows the variance of growth rate over time:
\begin{equation} \label{eq:sigma}
\sigma^2 = \sum_{i} p_i \, (\log \bar{f}_i - \mu)^2 \,.
\end{equation}
In this plot, different points represent different values of $q$ and their color represents the fixation probability. The maximum fixation probability is reached not at the peak, but on an ``efficient frontier'' to the left of the peak with smaller variance (it is not always better to have a smaller variance, as shown later). This demonstrates that fluctuations in the growth rate also plays a role in determining the outcome of competition in fluctuating environments. We will systematically study the effect of different types of fluctuations below.

\section{Stochastic contributions to extinction risk}

We expect all sources of stochasticity in the population dynamics to contribute to the extinction risk of a species. Besides environmental fluctuation, one type of stochasticity that is often neglected when studying bet-hedging is demographic fluctuation. It comes from the fact that population sizes are discrete numbers, changing in time due to stochastic events of births and deaths. The demographic rates of these events describe the expected changes of population sizes, but an actual trajectory of such stochastic processes will fluctuate over time and deviate from the average behavior. In calculating the long-term growth rate of the bet-hedging species, one works in the limit of effectively infinite population size where demographic fluctuation is negligible. But in a competition where population sizes are limited, demographic fluctuation can have significant effects as it causes small populations to go extinct by chance.

Another type of stochasticity that is unique to bet-hedging species is the fluctuation of phenotype composition within the population. Because the bet-hedging species diversifies into multiple phenotypes, the mean fitness of the population depends on the phenotype composition. However, even for a given probability $q$ with which each individual takes on a phenotype, the actual composition of the population may deviate from this probability because of finite population size. This is similar to the sampling noise one encounters when drawing a finite sample from a multinomial distribution. As the phenotypes are randomly drawn in every generation, the composition will fluctuate over time. Such compositional fluctuation means that, even if there is an optimal phenotype distribution for the bet-hedging species, it cannot be strictly followed at all times. In particular, when the population is small, the deviations will be significant (as we cannot have a ``chimeric'' individual that is partly one phenotype and partly another). This compositional fluctuation acts differently from demographic fluctuation, as it affects the mean fitness of the population, whereas demographic fluctuation causes randomness in selection given the fitness values.

In order to understand how these different types of fluctuations affect the survival of the bet-hedging species, we will examine their contributions separately. To do so, we will study the competition between different pairs of species, each pair chosen to illustrate a particular contribution. To isolate each contribution, we will approximate a bet-hedging species by a species with specific properties to remove other types of fluctuations as follows.

To remove compositional fluctuation, we calculate the mean fitness of the population according to its expected phenotype distribution $q$, i.e., $\bar{f}_i \equiv \sum_j f_{ij} \, q_j$. We then replace the original bet-hedging species with a homogeneous population that has fitness $\bar{f}_i$ in each environment $\varepsilon_i$. Note that the fitness of this substitute species still varies with the environment. Since the original bet-hedging species has reduced temporal variation of the mean fitness $\bar{f}_i$ by diversifying into multiple phenotypes, the substitute species will still enjoy such reduced fitness variation. The latter can be thought of as using a so-called ``conservative bet-hedging'' (CBH) strategy, as opposed to the original ``diversified bet-hedging'' (DBH) strategy \cite{Seger1987, philippi1989hedging}. Thus, we effectively approximate a DBH species by a CBH species to remove compositional fluctuation.

We can further remove environmental fluctuation by replacing the CBH species with a species that has a constant fitness value (CFV) regardless of the environmental state. To do so, we calculate the long-term growth rate $\mu$ of the CBH species according to Eq.~(\ref{eq:mu}), then use $\mathrm{e}^{\mu}$ (which is the geometric mean fitness) as the fitness value for the CFV species. This removes the fitness variation over time due to environmental fluctuation.

In the following, we will use different combinations of the DBH, CBH, and CFV species to study the stochastic contributions from each of the three types of fluctuations (Table~\ref{tab:cases}).

\begin{table}
\renewcommand{\arraystretch}{1.5}
\centering
\begin{tabular}{ |c|c|c|c| }
\hline
Fluctuations & Species 1 & Species 2  & Figures \\
\hline
demographic & CFV ($\mu_1(q)$ varies, $\sigma_1 = 0$) & CFV ($\mu_2 = 0$, $\sigma_2 = 0$) & Figure~\ref{fig:demo} \\
\hline
environmental & CBH ($\mu_1(q)$ varies, $\sigma_1(q) > 0$) & CFV ($\mu_2 = \mu_1$, $\sigma_2 = 0$) & Figure~\ref{fig:env} \\
\hline
compositional & DBH ($\mu_1(q)$ varies, $\sigma_1(q) > 0$) & CBH ($\mu_2 = \mu_1$, $\sigma_2 = \sigma_1$) & Figure~\ref{fig:compo} \\
\hline
combined & DBH ($\mu_1(q)$ varies, $\sigma_1(q) > 0$) & CFV ($\mu_2 = 0$, $\sigma_2 = 0$) & Figure~\ref{fig:EF} \\
\hline
\end{tabular}
\caption{Combinations of species models used to study different types of stochastic fluctuations.}
\label{tab:cases}
\end{table}

\subsection{Demographic fluctuation}

We start from the simplest case with only demographic fluctuation. This means both species will have constant fitness values, i.e., both will be CFV species. In our model, the demographic fluctuation comes from stochasticity in the Wright-Fisher selection process, according to which every new generation is randomly resampled from the previous generation. 

This case is a classic problem studied in population genetics. Let the two species have fitness values $\mathrm{e}^{\mu_1}$ and $\mathrm{e}^{\mu_2}$. When there is no fitness difference between the two species (i.e., $\mu_1 = \mu_2$, the neutral case), their fixation probability will be equal to their initial fraction in the total population. Let $s = \mu_1 - \mu_2$, then for $s \neq 0$, the fixation probability of species 1 is given by the well-known result \cite{kimura1962probability}
\begin{equation} \label{eq:pfix}
\theta = \frac{1 - \mathrm{e}^{-2 s N x}}{1 - \mathrm{e}^{-2 s N}} \,,
\end{equation}
where $x$ is the initial fraction of species 1 in the total population.

\begin{figure}
\centering
\includegraphics[width=0.5\textwidth]{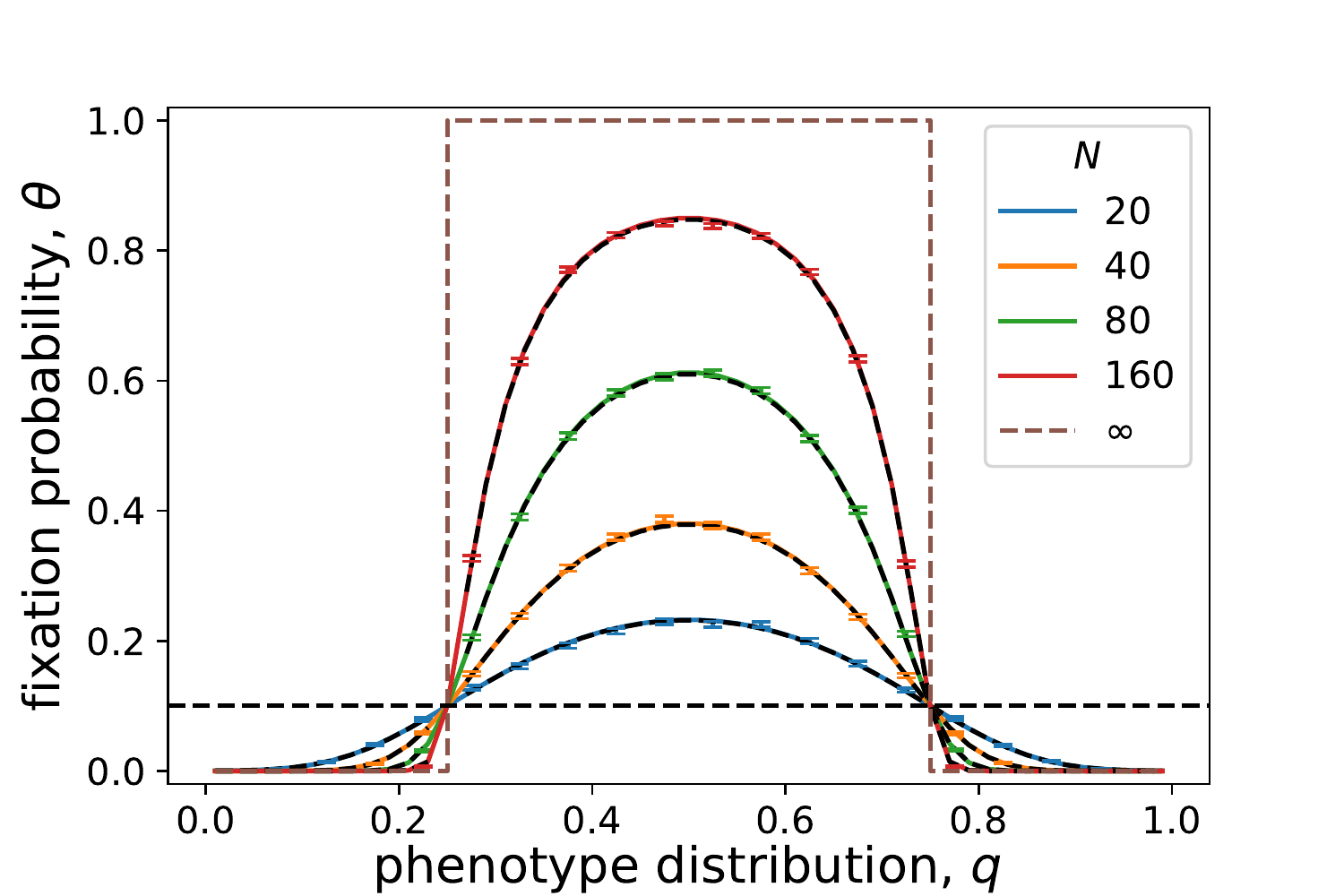}
\caption{\small Effect of demographic fluctuation on fixation probability $\theta$. We compete two CFV species with $\sigma_1 = \sigma_2 = 0$; $\mu_1$ is a function of phenotype distribution $q$ according to Eq.~(\ref{eq:mu}), and $\mu_2$ is that for $q = 0.75$. Solid curves are numerical results for different total population sizes $N$; points with errorbars are from simulations; dashed curves are theoretical results from Eq.~(\ref{eq:pfix}). The black horizontal dashed line marks the initial fraction of species 1, $x = 0.1$, which represents the fixation probability in the neutral case. The brown dashed line represents the limit of infinite population size, for which demographic fluctuation vanishes. For smaller $N$, demographic fluctuation pulls the curves towards the neutral case, thus hurting the fitter species.}
\label{fig:demo}
\end{figure}

In our simulations we keep $\mu_2$ constant while varying $\mu_1$ as a function of $q$ according to Eq.~(\ref{eq:mu}). Figure~\ref{fig:demo} shows the fixation probability as a function of the bet-hedging ratio $q$ for different total population sizes. Species 1 initially forms a fraction $x = 0.1$ of the total population in all these simulations (see Section~\ref{sec:initial} for other cases). Given our parameter values, we have $s > 0$ for $0.25 < q < 0.75$, and $s < 0$ otherwise. When the total population size $N$ goes to infinity, the fixation probability becomes $1$ for $s > 0$ and $0$ for $s < 0$, i.e., the fitter species always wins. For a finite population size, Figure~\ref{fig:demo} shows that the curves are smooth and approach the neural case. Thus, demographic fluctuation reduces the fixation probability of the fitter species. The effect is more significant for small population sizes (see Section~\ref{sec:total} for more discussion). Our numerical and simulation results both agree perfectly with the theoretical formula above.

\subsection{Environmental fluctuation} \label{sec:enviro}

To study the effect of environmental fluctuation, we compete a CBH species with a CFV species, whose long-term growth rates are equal. Both species are subject to demographic fluctuation but not compositional fluctuation; the only difference between them is that the CBH species is also subject to environmental fluctuation that causes its fitness to vary over time. In this way we can study how the environmental fluctuation will affect the survival of the CBH species. As we vary the probability $q$, both the long-term growth rate $\mu_1$ and variance $\sigma_1$ of the CBH species vary according to Eqs.~(\ref{eq:mu}--\ref{eq:sigma}), while for the CFV species $\mu_2$ is always set equal to $\mu_1$ and $\sigma_2 = 0$.

\begin{figure}
\centering
\includegraphics[width=0.5\textwidth]{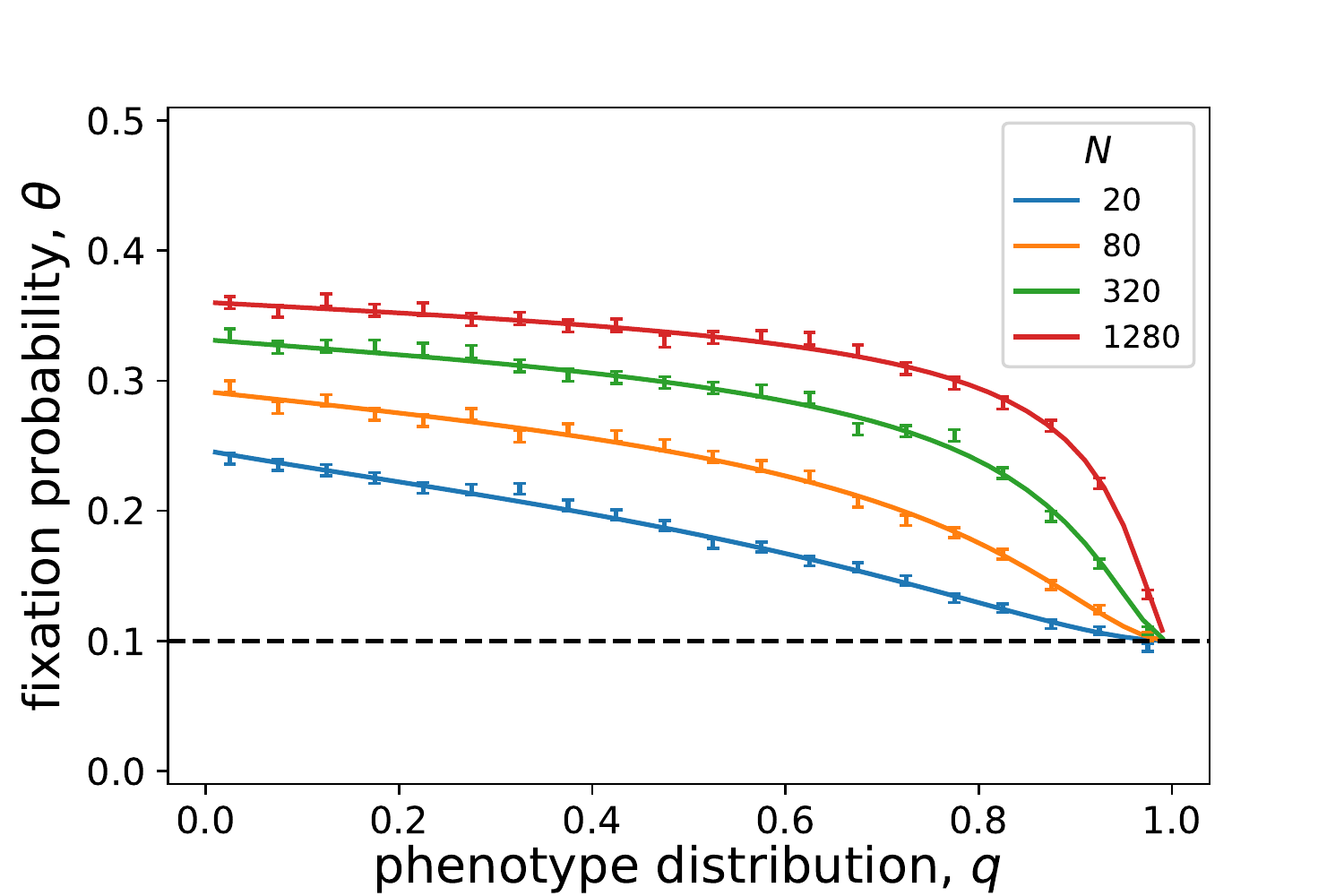}
\caption{\small Effect of environmental fluctuation on fixation probability. We consider a CBH species competing with a CFV species, with $\mu_1$ and $\sigma_1$ dependent on $q$ according to Eqs.~(\ref{eq:mu}--\ref{eq:sigma}) while $\mu_2$ is set equal to $\mu_1$ and $\sigma_2 = 0$. The black dashed line is the initial fraction of the CBH species, $x = 0.1$, which represents the neutral case without environmental fluctuation. Solid curves are numerical results; points with errorbars are from simulations. The fixation probability is higher than the initial fraction, which means that environmental fluctuation enhances the survival of the CBH species.}
\label{fig:env}
\end{figure}

The fixation probability $\theta$ of the CBH species as a function of $q$ is shown in Figure~\ref{fig:env}. For $q = 1$, the variance $\sigma_1$ vanishes because all individuals have the dormant phenotype that has the same fitness in both environments. In this case, there is no environmental fluctuation for both species, hence the model reduces to the previous case with only demographic fluctuation. Since $\mu_1 = \mu_2$ by construction, this is the neutral case where the fixation probability $\theta$ is equal to the initial fraction $x$, as confirmed in Figure~\ref{fig:env}.

For $q < 1$, whereby $\sigma_1 > 0$, Figure~\ref{fig:env} shows that the fixation probability is higher than the initial fraction. In other words, environmental fluctuation appears to increase the survival probability of the species. This seems to contradict the notion of a mean-variance trade-off (Figure~\ref{fig:EF} inset) that is common in economics and ecology \cite{Laureti2009} --- One might have expected that, if the mean (i.e., geometric mean fitness, or the long-term growth rate) is fixed, then the variance would have contributed to more stochasticity in the population growth and thus increase the extinction risk. Instead, we find that environmental fluctuation can actually reduce the extinction risk. This is true as long as the species is initially a small fraction of the total population (see Section~\ref{sec:initial}, Figure~\ref{fig:initial-fraction}).

\subsection{Compositional fluctuation} \label{sec:compo}

The compositional fluctuation is unique to the DBH species, arising only because the population diversifies into different phenotypes. To study the effect of this unique type of fluctuation, we compete a DBH species with a CBH species while keeping the geometric mean and variance of their fitness the same, i.e., $\mu_1 = \mu_2$ and $\sigma_1 = \sigma_2$. Both species are subject to the same demographic and environmental fluctuations; their only difference is that the DBH species has a phenotype composition that fluctuates over time, whereas the CBH species is always homogeneous.

If the phenotype composition of the DBH species were precisely equal to $q$ (which can only happen when its population size approaches infinity), then its population mean fitness would be equal to the CBH species. In this case, even though the fitness of both species vary over time as the environment fluctuates, they would be neutral with respect to each other in every generation. Hence the fixation probability $\theta$ would again be equal to the initial fraction $x$. In contrast, when the phenotype composition of the DBH species is allowed to fluctuate, any deviation of its fixation probability from the initial fraction will represent the effect of compositional fluctuation.

\begin{figure}
\centering
\includegraphics[width=0.5\textwidth]{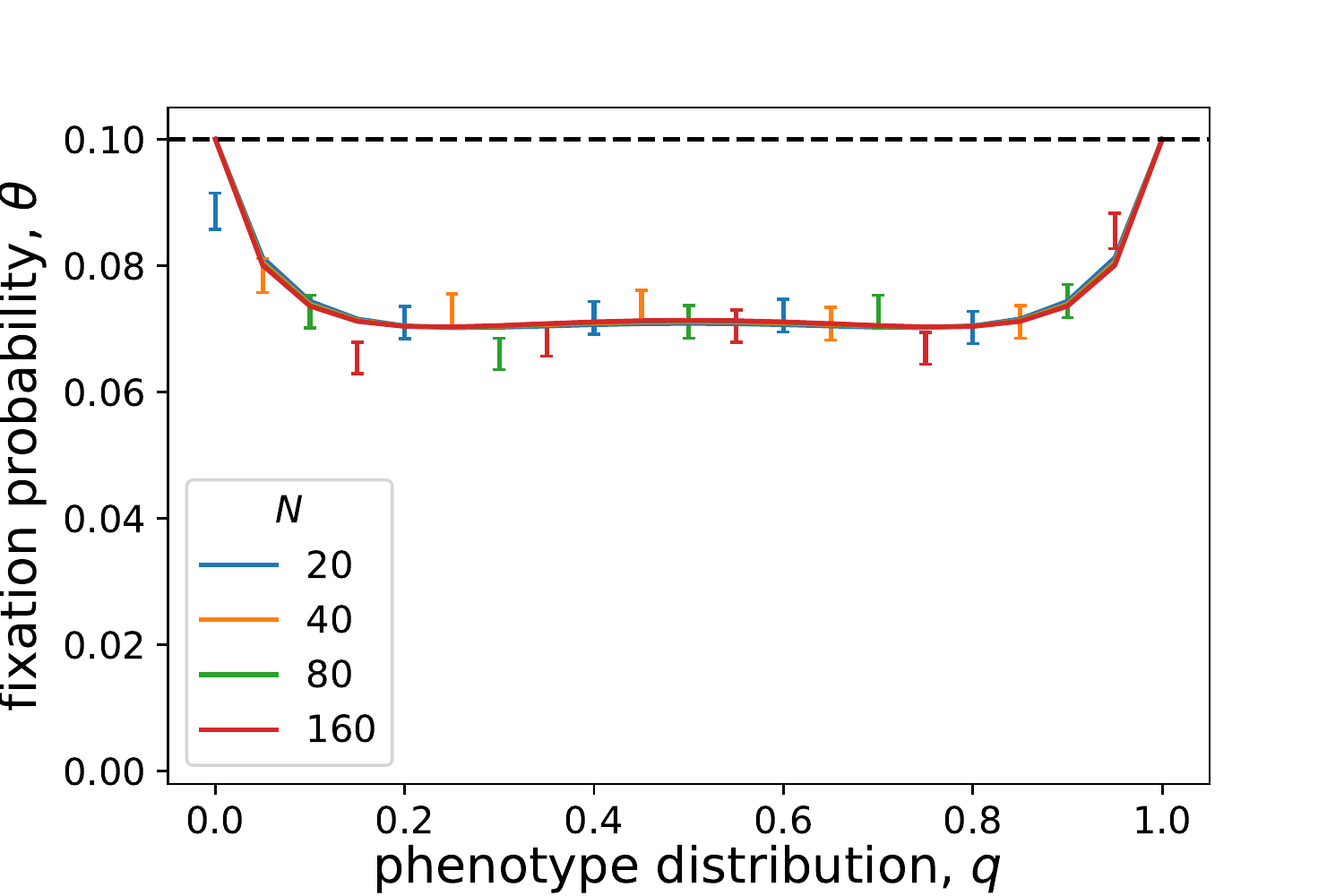}
\caption{\small Effect of compositional fluctuation on fixation probability. We consider a DBH species competing with a CBH species, with $\mu_1 = \mu_2$ and $\sigma_1 = \sigma_2$, except that the DBH species has diverse phenotypes whereas the CBH species is homogeneous. The black dashed line is the initial fraction of the DBH species, $x = 0.1$, which represents the result when there is no compositional fluctuation. Solid curves are numerical results; points with errorbars are from simulations. Compositional fluctuation decreases the fixation probability and its effect does not depend on the total population size $N$.}
\label{fig:compo}
\end{figure}

Figure~\ref{fig:compo} shows how the fixation probability $\theta$ depends on $q$. It can be seen that the fixation probability is smaller than the initial fraction, $x = 0.1$. Thus, compositional fluctuation always increases the extinction risk. Since the variance of the phenotype composition is proportional to $q (1-q)$, the effect of compositional fluctuation vanishes when $q$ is close to $0$ or $1$. Notably, the effect of compositional fluctuation does not depend on the total population size $N$, in contrast to demographic and environmental fluctuations. We will study the differences between these stochastic contributions in the next section.

\section{Comparison of stochastic contributions} \label{sec:discussion}

A bet-hedging species would experience all three types of fluctuations illustrated above. To characterize how their contributions to the extinction risk differ, we will study how their effects depend on the parameters of the model.

\subsection{total population} \label{sec:total}

\begin{figure}
\centering
\includegraphics[width=0.5\textwidth]{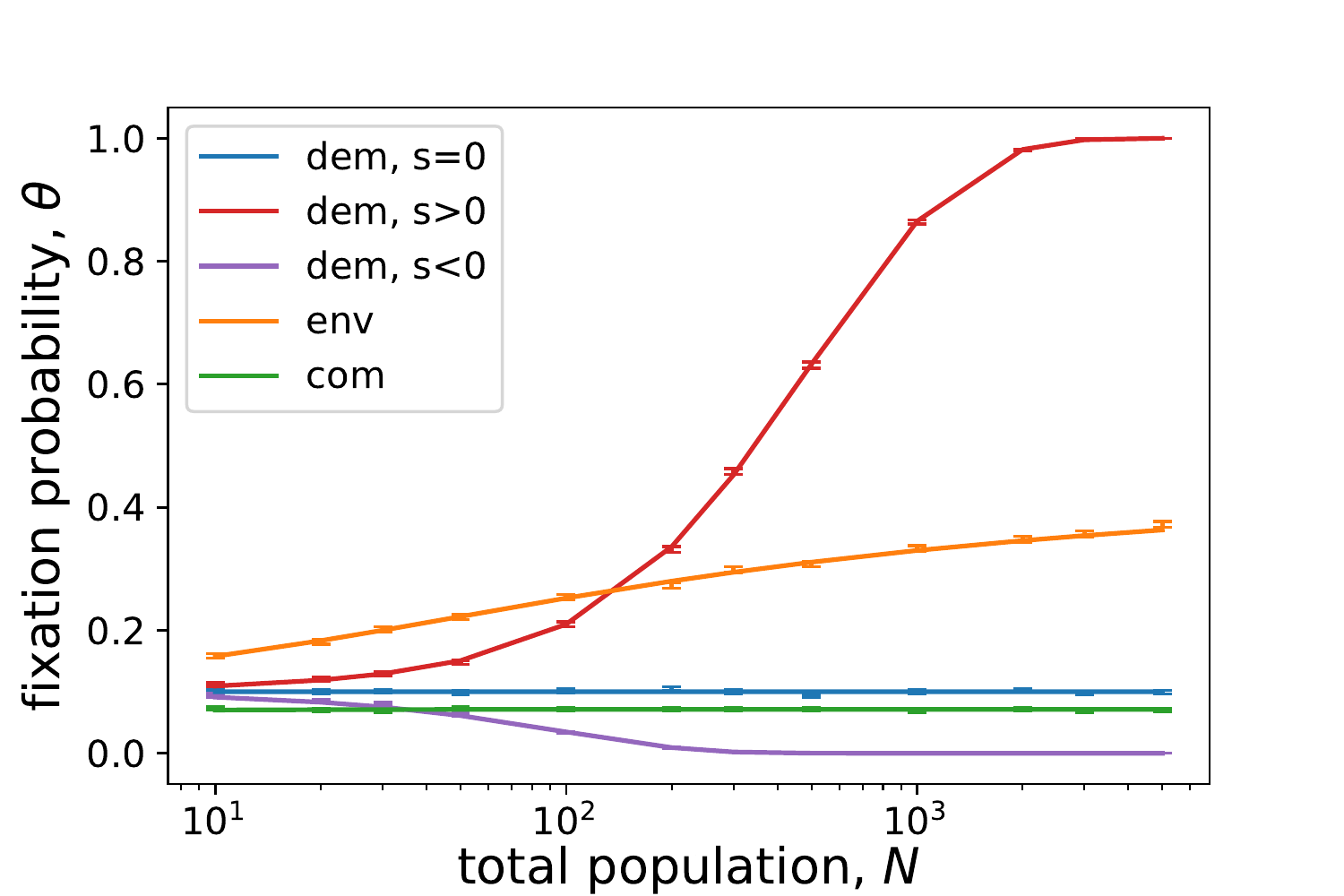}
\caption{\small Dependence of fixation probability $\theta$ on the total population size $N$. Solid curves are numerical results for the fixation probability of species 1 ($q = 0.5$) under different types of fluctuations (Table~\ref{tab:cases}); points with errorbars are from simulations. The blue line is the neutral case where $\theta$ is equal to the initial fraction, $x = 0.1$. Demographic fluctuation ($s = \pm 0.01$) significantly reduces the advantage of the fitter species when $N$ is small. Environmental fluctuation has a stronger effect in helping species survive when $N$ is large. Compositional fluctuation has a negative effect on the species survival that does not depend on $N$.}
\label{fig:total-population}
\end{figure}

The compositional fluctuation has not been analyzed before partly because it did not show up in previous studies that focused on the long-term growth rate of the bet-hedging species, which treated the population size as infinite. One might then expect that, if the total population size $N$ in our model becomes large, the effect of compositional fluctuation would disappear. However, this is not the case. As shown in Figure~\ref{fig:compo} and further illustrated in Figure~\ref{fig:total-population}, the fixation probability corresponding to compositional fluctuation does not depend on $N$. This can be understood by noting that the two species we used for studying compositional fluctuation have $\mu_1 = \mu_2$, like in the neutral case ($s = 0$) for demographic fluctuation. In the latter case, the fixation probability also does not depend on $N$. Nevertheless, for $s > 0$, the fixation probability decreases significantly in the small $N$ limit, showing that demographic fluctuation \emph{hurts the fitter} species. In contrast, the effect of environmental fluctuation is more significant for large $N$.

\subsection{initial fraction} \label{sec:initial}

\begin{figure}
\centering
\includegraphics[width=0.5\textwidth]{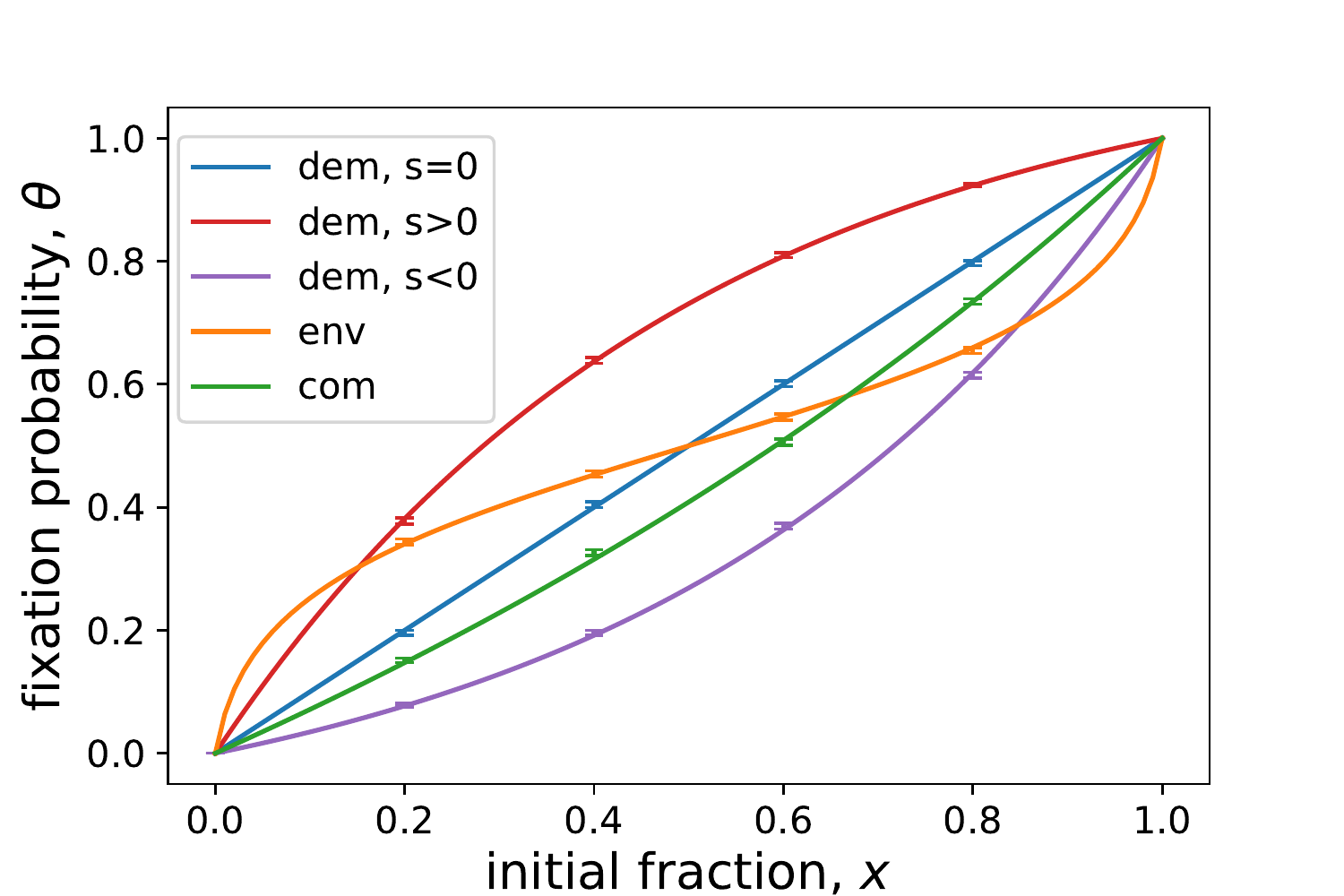}
\caption{\small Dependence of fixation probability $\theta$ of species 1 ($q = 0.5$) on its initial fraction $x$ in the total population ($N = 100$). Solid curves are numerical results for different types of fluctuations; points with errorbars are from simulations. The diagonal line (blue) represents the neutral case. Environmental fluctuation has a positive effect when $x < 0.5$ but a negative effect when $x > 0.5$, which means it helps the rarer species survive and thus promotes coexistence. Compositional fluctuation always has a negative effect on survival.}
\label{fig:initial-fraction}
\end{figure}

So far we have considered the case where the bet-hedging species is initially a small fraction ($x = 0.1$) of the total population. Our results may be relevant for the situation where a bet-hedging species tries to invade another species, such that the initial population of the invading species would be much less than the resident species. If we vary the initial fraction, the effects of different fluctuations are shown in Figure~\ref{fig:initial-fraction}. As a reference, a straight diagonal line where the fixation probability $\theta$ is equal to the initial fraction $x$ represents the neutral case ($s = 0$). The curves for demographic fluctuation with $s > 0$ and $s < 0$ are fully above and below the diagonal, respectively. In contrast, the curve for environmental fluctuation crosses the diagonal at $x = 0.5$, showing that environmental fluctuation increases the fixation probability of a species only when it is relatively rare ($x < 0.5$). Similar results have been obtained in previous models, e.g., \cite{melbinger2015impact, Dean2020}. Note that, if this species becomes the majority of the total population ($x > 0.5$), its fixation probability will then be reduced by environmental fluctuation, meaning that the other species, now the rarer one, will have an increased fixation probability. Therefore, environmental fluctuation \emph{benefits the smaller} population, and thus promotes coexistence. As for compositional fluctuation, the curve is fully below the diagonal, showing that it always reduces the fixation probability.

\subsection{overlapping generations} \label{sec:overlap}

We can generalize our model to include overlapping generations, such that only a fraction of the population dies every generation. This can be modeled by introducing a parameter $d$ that represents the probability that each individual will be replaced by new individuals at every time step (see Appendix for mathematical details). Figure~\ref{fig:overlapping-generations} shows the fixation probability as a function of $d$ for all three types of fluctuations. Our results above for non-overlapping generations correspond to $d = 1$. A smaller $d$ means fewer individuals are replaced every generation, which implies a smaller effective population size and hence a stronger effect of demographic fluctuation. This can be seen in the figure as the $s \neq 0$ curves both go towards the neutral limit when $d$ decreases. In contrast, the effect of compositional fluctuation becomes weaker for smaller $d$. That is because the phenotype composition of the population fluctuates less when fewer individuals are born every generation with randomly assigned phenotypes. Indeed, the curve for compositional fluctuation goes up slightly for smaller $d$, showing that it has a \emph{weaker} effect in reducing the fixation probability. On the other hand, the fixation probability under environmental fluctuation increases with smaller $d$, implying a stronger effect in promoting species survival when generations overlap.

\begin{figure}
\centering
\includegraphics[width=0.5\textwidth]{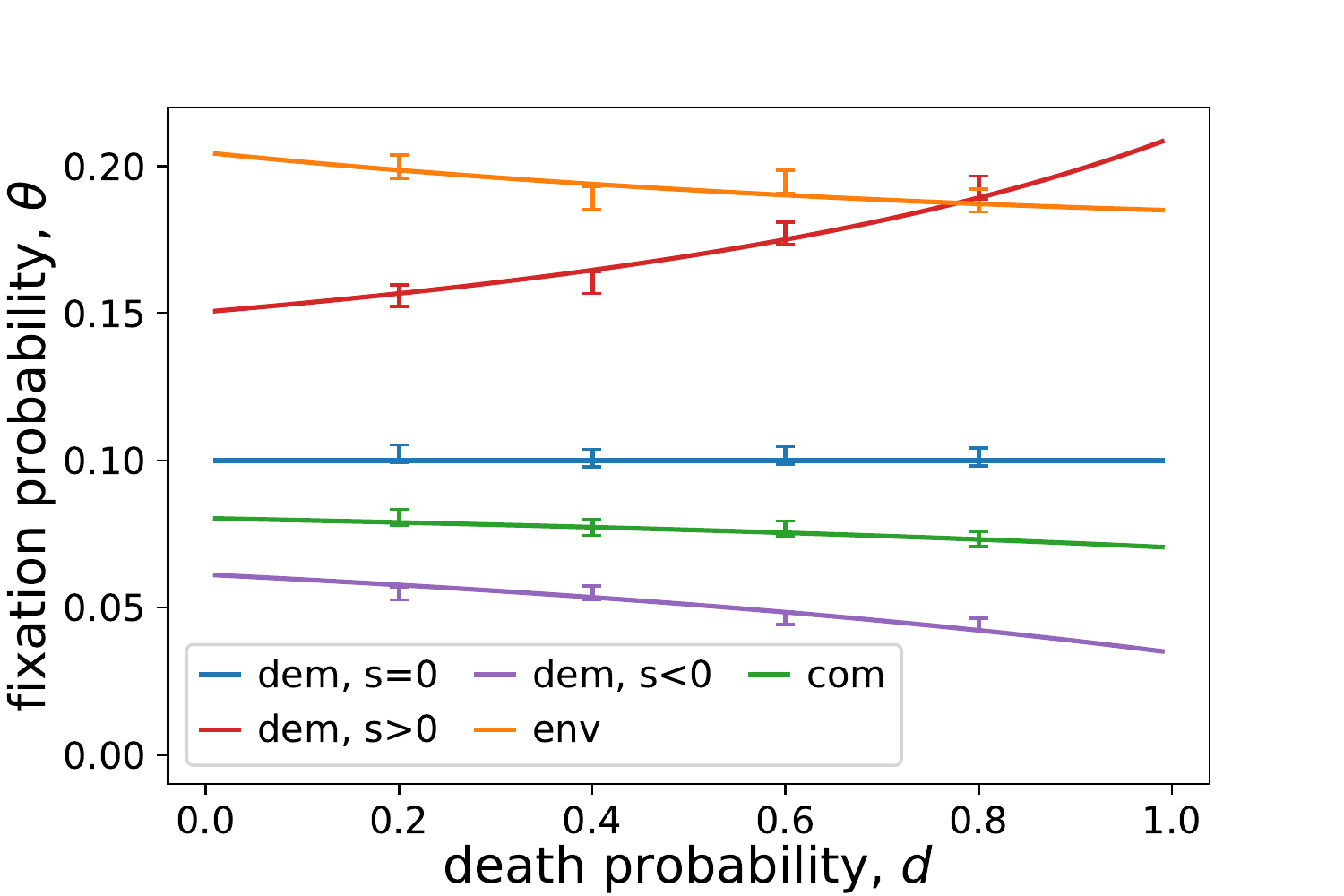}
\caption{\small Dependence of fixation probability $\theta$ on the death probability $d$ for overlapping generations. The phenotype distribution of species 1 is $q = 0.8$, with an initial fraction $x = 0.1$, and total population size $N = 100$. Solid curves are numerical results; points with errorbars are from simulations. The blue line represents the neutral case without environmental and compositional fluctuations.  Demographic and environmental fluctuations have a stronger effect for smaller $d$, i.e., more overlapping generations, whereas compositional fluctuation has a slightly weaker effect.}
\label{fig:overlapping-generations}
\end{figure}

The idea that environmental fluctuation can help species survive has been studied among ``fluctuation-dependent coexistence mechanisms'' in ecology \cite{chessen:2000, Chesson2018, Dean2020}. One such mechanism is the ``storage effect'', which can be explained by a heuristic argument as follows \cite{chesson1981environmental}. In each generation, on average, $d N$ individuals will die and be replaced. Among these, a fraction of $P = \frac{f_1 N_1}{f_1 N_1 + f_2 N_2}$ will be replaced by new individuals of the first species, where $N_1$ and $N_2$ are the current population sizes of the two species and their fitness $f_1$ and $f_2$ may depend on time. Therefore, in the next generation the expected population size of the first species will be $N_1' = (1-d) N_1 + d N P$. When this species is at low density, i.e., $N_1 \ll N_2$, its population growth rate will be
\begin{equation} \label{eq:storage}
r_1 \equiv \log \frac{N_1'}{N_1} \approx \log \big( 1 - d + d \, \mathrm{e}^{s} \big) \,,
\end{equation}
where $s = \log (f_1 / f_2)$, which varies over time due to environmental fluctuation. The two species are assumed to have the same geometric mean fitness, so that $\mathbb{E}[s] = \mu_1 - \mu_2 = 0$. When $d < 1$, $r_1$ is a \emph{convex} function of $s$ around 0. Therefore, by Jensen's inequality, we have $\mathbb{E}[r_1(s)] > r_1(\mathbb{E}[s]) = 0$. This means that the first species has a positive average growth rate when its population is small, hence it tends to avoid extinction. Note that by this argument, the effect would disappear when $d = 1$, i.e., for non-overlapping generations. Our results are different as we focus on the fixation probability instead of the low-density growth rate, and in our case environmental fluctuation reduces extinction risk even in the limit $d \to 1$.

We can understand the effect of compositional fluctuation using a similar argument. In this case, the low-density growth rate of the first species (DBH) is given by the same Eq.~(\ref{eq:storage}), except that $f_1$ is replaced by the population mean fitness $\bar{f}_1$ that depends on the phenotype composition. Thus, we can write
\begin{equation}
r_1 \approx \log \big( 1 - d + d \, u \big) \,,
\end{equation}
where $u = \bar{f}_1 / f_2$, which fluctuates over time because the actual phenotype composition deviates from the expected value $q$. Since the second species (CBH) is chosen to have fitness $f_2$ given by the expected composition $q$, we have $\mathbb{E}[u] = 1$. Now, $r_1$ is a \emph{concave} function of $u$, hence $\mathbb{E}[r_1(u)] < r_1(\mathbb{E}[u]) = 0$. Therefore, compositional fluctuation reduces the average growth rate of the diversified species when its population is small, thus making it more likely to go extinct. Compared to the storage effect, here the inequality sign is reversed. As a result, while environmental fluctuation helps a small population survive, compositional fluctuation has the opposite effect of increasing the extinction risk. This effect is weaker for overlapping generations partly because the function $r_1(u)$ is less concave for smaller $d$.

\subsection{environmental correlation} \label{sec:correlation}

\begin{figure}
\centering
\includegraphics[width=0.5\textwidth]{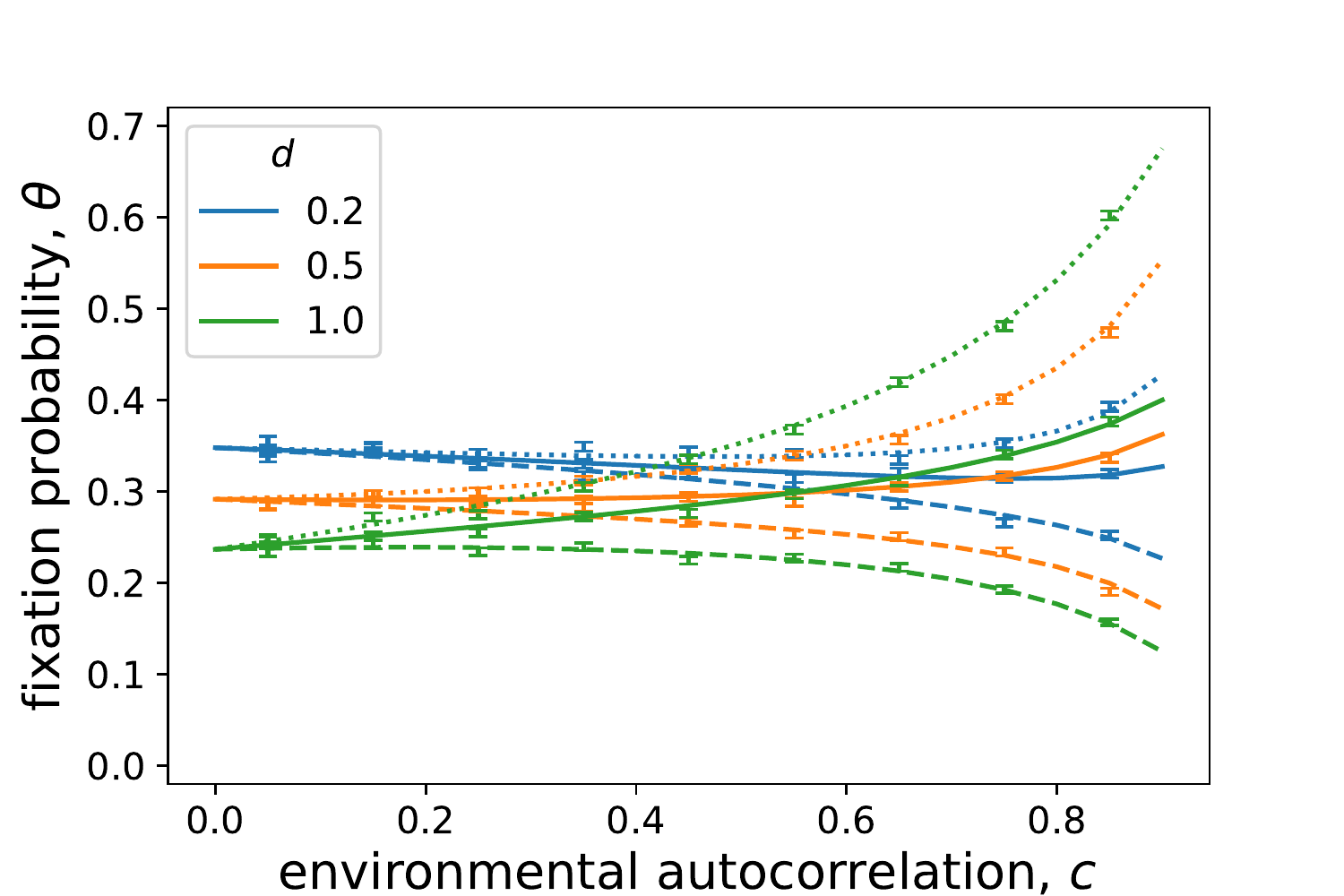}
\caption{\small Dependence of fixation probability $\theta$ on the environmental autocorrelation $c$. Whereas demographic and compositional fluctuations do not depend on $c$ (not plotted), environmental fluctuation does and, for large $c$, the resulting fixation probability also depends on the initial environmental state. The dashed and dotted curves are numerical results for initial state $\varepsilon_X$ and $\varepsilon_Y$, respectively. Solid curves are the average over the initial states; points with errorbars are from simulations. Other parameters are: total population size $N = 100$, initial fraction $x = 0.1$, and phenotype distribution $q = 0.6$.}
\label{fig:env-correlation}
\end{figure}

In our model so far, the environmental state at every time step is drawn independently from a probability distribution. We now consider the case where the environmental states are correlated over time. At every time step, we assume that the environment changes with probability $(1-c)/2$, where $c$ is the amount of auto-correlation. Varying $c$ does not affect demographic and compositional fluctuations because our models are set up to isolate these contributions. Therefore we will focus here on the fixation probability under environmental fluctuation only, as shown in Figure~\ref{fig:env-correlation}. As the auto-correlation $c$ increases from zero, the fixation probability starts to depend on the initial environmental state. This is due to the increase of the correlation time of the environment --- a longer correlation time means each environmental state lasts longer before switching, hence the influence of the initial state is not immediately lost. If the correlation time exceeds the mean extinction time of the species (see Appendix Figure~\ref{fig:mean-ext}), then the environment would not have switched before one species has likely gone extinct. In that case, the buffering of temporal fitness variation through bet-hedging is not utilized. Therefore, we have focused on short correlation times ($c \to 0$) in our analyses above. For overlapping generations ($d < 1$), the generation time and hence extinction time is longer, so the fixation probability depends less on the initial environmental state.

\section{Conclusion}

We have analyzed the effect of three distinct types of stochastic fluctuations on the extinction risk of a bet-hedging species when competing with other species. In summary, demographic fluctuation hinders the fitter species, environmental fluctuation favors the rarer species, and compositional fluctuation hinders the heterogeneous species (by ``hinder'' or ``favor'' we mean ``increase or decrease the extinction risk of'').

Bet-hedging is often considered as a beneficial strategy for surviving in fluctuating environments. The basic idea has been that, by diversifying into multiple phenotypes, the species reduces temporal variation of population mean fitness, thus becoming less affected by environmental changes. However, our results challenge this idea in the situation where the bet-hedging species has to compete with other species. For example, when a bet-hedging species invades another species, temporal variation of fitness could be beneficial due to fluctuation-dependent coexistence mechanisms, such as the storage effect. Thus, it is not necessarily best for the bet-hedging species to minimize the fitness variation caused by environmental fluctuation. Moreover, by diversifying into multiple phenotypes, the bet-hedging species incurs an additional type of stochasticity in population dynamics --- the compositional fluctuation --- that tends to lower its chance of survival. Our results suggest that bet-hedging may not be as beneficial as previously considered. Beyond the long-term population growth rate often used as the standard for evaluating the benefit of this strategy, other factors such as the extinction risk in competitive situations may be important for determining its significance in promoting species survival and coexistence in varying environments.

\section*{Acknowledgements}

We thank Nicholas Kortessis for extremely helpful discussions and comments.

\appendix

\setcounter{figure}{0}
\renewcommand\thefigure{A\arabic{figure}}
\setcounter{equation}{0}
\renewcommand\theequation{A\arabic{equation}}

\section{Simulation methods}

We study a model with two environmental states ($\varepsilon_X$ and $\varepsilon_Y$) and two phenotypes ($\phi_A$ and $\phi_B$). We choose the environment distribution to be $(p_A, p_B) = (0.5, 0.5)$, which means the probabilities of the two environmental states are equal. In the case where the environment is uncorrelated over time, we draw an environmental state randomly at every time step. When there is auto-correlation, at every time step the environment has a probability $(1-c)/2$ to switch and a probability $(1+c)/2$ to remain in the same state, where $c$ is the amount of auto-correlation.

For the bet-hedging species, each individual has a probability $q$ of having phenotype $\phi_A$ and $1-q$ of having $\phi_B$. At every time step, the number of phenotype $\phi_A$ within species 1, denoted by $N_{1A}$, is drawn from a binomial distribution $B(N_1,q)$ where $N_1$ is the population size of species 1; then $N_{1B} = N_1 - N_{1A}$. In a given environment, we determine the fitness of each phenotype according to the fitness matrix $f_{ij} \equiv f(\varepsilon_{i},\phi_{j}) = \left( \begin{smallmatrix} f_{XA} & f_{XB} \\ f_{YA} & f_{YB} \end{smallmatrix} \right)$.

We use the Wright-Fisher model to describe the selection process at every generation. Suppose the current population sizes of the two species are $N_1$ and $N_2$, and the current environmental state is $\varepsilon_i$. The population size $N_1'$ of species 1 in the next generation will be drawn from a binomial distribution $B(N,P_i)$ with probability $P_i = N_1 \bar{f}_{1,i} / (N_1 \bar{f}_{1,i} + N_{2} f_{2,i})$, where $\bar{f}_{1,i} = (N_{1A} f_{iA} + N_{1B} f_{iB}) / N_1$ is the mean fitness of species 1 in the environment $\varepsilon_i$ and $f_{2,i}$ is the fitness of species 2 (in the cases we studied, listed in Table~\ref{tab:cases}, species 2 does not diversify). Since the total population size is fixed, we have for the second species $N_2' = N - N_1'$. When we consider overlapping generations, only an average fraction $d$ of individuals is replaced at each time step. The actual number of species 1 that will be replaced, $R_1$, is drawn from a binomial distribution $B(N_1,d)$. Similarly, for species 2 we draw $R_2$ from a distribution $B(N_2,d)$. From these we get the total number of individuals to be replaced in that time step, $R = R_1 + R_2$. Then we replace them by drawing a number $R_1'$ from a binomial distribution $B(R,P_i)$, and $R_2' = R - R_1'$. In the next time step, we have $N_1' = N_1 - R_1 + R_1'$ and $N_2' = N - N_1'$. The original Wright-Fisher model corresponds to $d = 1$.

In this way, we update the population size of each species at every time step. When one of the species reaches zero population size, this species is extinct and the other one is fixed, terminating the simulation. We run the simulation 10000 times and count how often species 1 reaches fixation. This fixation probability is used to evaluate the performance of the bet-hedging species.

\section{Numerical calculation}

The fixation probability can also be calculated numerically. To do that, we write down the transition probability matrix for both the environmental state and the population size,
\begin{equation}
    \mathbb{P}(N_1', \varepsilon_k | N_1, \varepsilon_i) = b(N_1';N,P_i) \, p(\varepsilon_k) \,,
\end{equation}
where $b(N_1';N,P_i)$ is the probability mass function for the binomial distribution $B(N,P_i)$ used above for simulations, and $p(\varepsilon_k)$ is the probability of the environmental state $\varepsilon_k$. If there is environmental correlation, then $p(\varepsilon_k)$ is replaced by the switching probability $p(\varepsilon_k|\varepsilon_i)$. When considering overlapping generations, the binomial distribution $b(N_1';N,P_i)$ is replaced by
\begin{equation}
    \sum_{R_1, R_2} b(R_1; N_1, d) \, b(R_2; N_2, d) \, b(N_1'-N_1+R_1; R_1+R_2, P_i) \,.
\end{equation}

The transition probability matrix is a $2(N\!\!+\!\!1) \times 2(N\!\!+\!\!1)$ matrix that represents the selection process at a single time step. For the transition probability over $T$ time steps, we take the $T$-th power of this matrix. In the long term ($T \to \infty$), the matrix power converges to a stationary matrix $\mathbb{P}^\infty (N_1', \varepsilon_k | N_1, \varepsilon_i)$, which can be calculated using eigenvalue decomposition. From this we calculate
\begin{equation} \label{eq:trans-inf}
\mathbb{P}^\infty(N_1' | N_1) = \sum_{\varepsilon_k, \varepsilon_i} \mathbb{P}^\infty (N_1', \varepsilon_k | N_1, \varepsilon_i) \, p(\varepsilon_i) \,.
\end{equation}
The entries of this $(N\!\!+\!\!1) \times (N\!\!+\!\!1)$ matrix is $0$ except for $N_1' = 0$ or $N$. The fixation probability of species 1 starting from an initial population size $N_1$ is given by the entry $\mathbb{P}^\infty(N_1' \!=\! N | N_1)$.

\section{Extinction time distribution}

\begin{figure}
\centering
\includegraphics[width=0.5\textwidth]{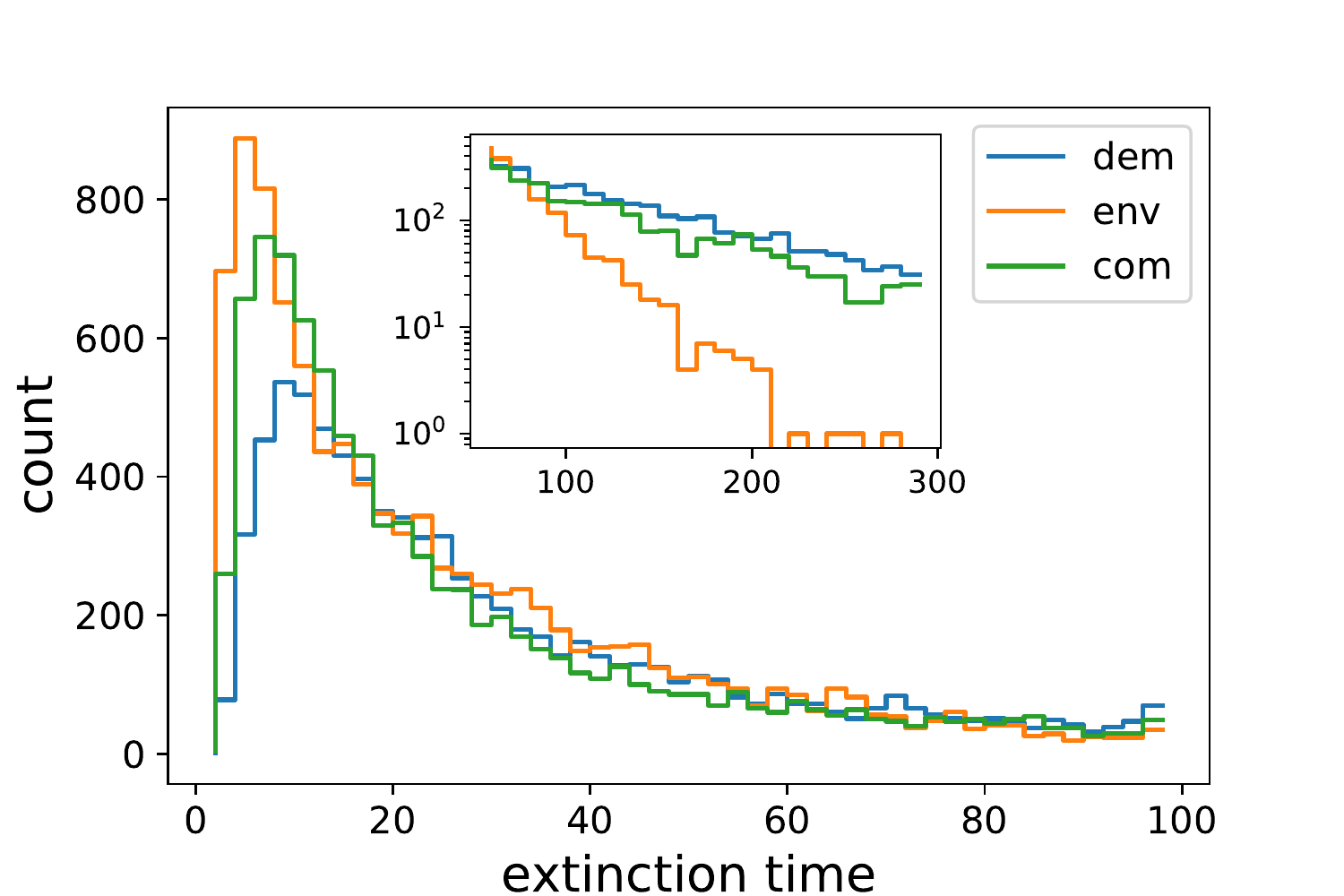}
\caption{\small Distributions of extinction time under three types of fluctuations studied in the main text. These distributions are obtained from simulations with total population size $N = 100$, initial fraction $x = 0.1$, and phenotype distribution $q = 0.6$ for the bet-hedging species.}
\label{fig:ext-time}
\end{figure}

Although it is difficult to calculate the extinction time distribution analytically, we can estimate it from simulations. We run the simulation 10000 times for each set of parameters and record the time when extinction happens in each run. From these we can plot the distribution of extinction times (Figure~\ref{fig:ext-time}) and calculate the mean extinction time.

\begin{figure}[h]
\centering
\includegraphics[width=0.5\textwidth]{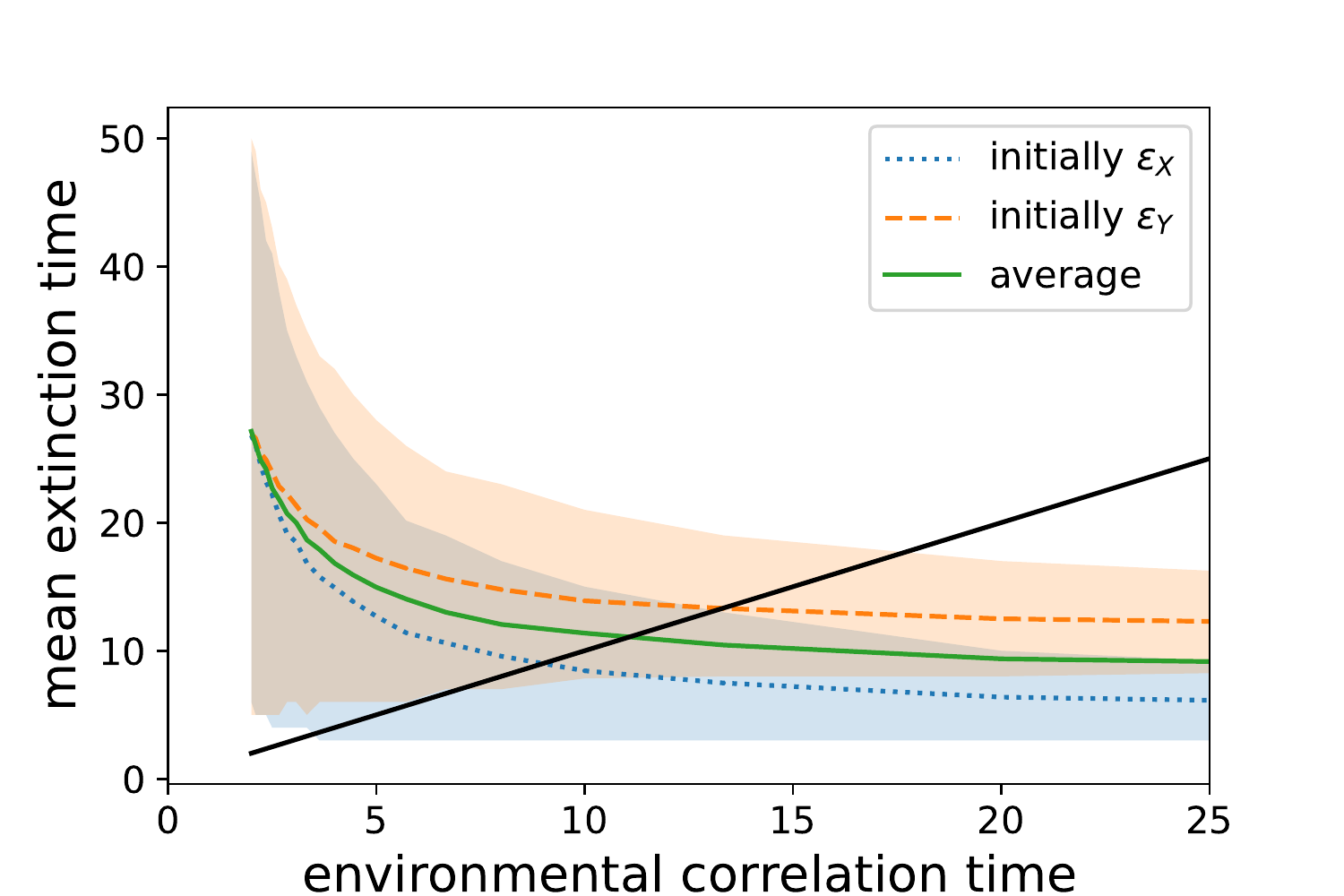}
\caption{\small Mean extinction time under environmental fluctuations with different correlation times. The results are for a CBH species competing with a CFV species, with $N = 100$, $x = 0.1$, and $q = 0.6$. The dashed and dotted curves are simulation results for initial environmental state $\varepsilon_X$ and $\varepsilon_Y$ respectively, and the solid curve is their average. Shaded area represents the 16--84 percentile range of the extinction time distribution. The black line is where the mean extinction time equals environmental correlation time. When environmental correlation time exceeds the mean extinction time, the environment would not have switched before one species has likely gone extinct.}
\label{fig:mean-ext}
\end{figure}

When the mean extinction time becomes comparable to the environmental correlation time (measured by the average duration of each environmental state, $\tau = \frac{2}{1-c}$ ), the fixation probability will depend on the initial environmental state. In that case, we cannot average over the initial environment as in Eq.~(\ref{eq:trans-inf}). Instead, we calculate
$\mathbb{P}^\infty(N_1' | N_1, \varepsilon_i)$ for each initial state $\varepsilon_i$ by summing over $\varepsilon_k$ only. Figure~\ref{fig:mean-ext} shows the mean extinction time for different $\varepsilon_i$, as well as their average. In this example, when the environmental correlation time exceeds about $10$ ($c \gtrsim 0.8$), it is no longer meaningful to study the effect of bet-hedging in changing environments, because the environment would not have changed before one species has likely gone extinct.

\clearpage
\bibliographystyle{unsrt}
\bibliography{bibliography}

\begin{thebibliography}{10}

\bibitem{Tilman2011}
David Tilman.
\newblock {Diversification, biotic interchange, and the universal trade-off
  hypothesis}.
\newblock {\em The American Naturalist}, 178(3):355--371, sep 2011.

\bibitem{levins1979coexistence}
Richard Levins.
\newblock Coexistence in a variable environment.
\newblock {\em The American Naturalist}, 114(6):765--783, 1979.

\bibitem{hutchinson:1961}
G.~E. Hutchinson.
\newblock The paradox of the plankton.
\newblock {\em The American Naturalist}, 95(882):137--145, 1961.

\bibitem{chesson1981environmental}
Peter~L Chesson and Robert~R Warner.
\newblock Environmental variability promotes coexistence in lottery competitive
  systems.
\newblock {\em The American Naturalist}, 117(6):923--943, 1981.

\bibitem{chessen:2000}
Peter Chesson.
\newblock Mechanisms of maintenance of species diversity.
\newblock {\em Annual Review of Ecology and Systematics}, 31(1):343--366, 2000.

\bibitem{Chesson2018}
Peter Chesson.
\newblock {Updates on mechanisms of maintenance of species diversity}.
\newblock {\em Journal of Ecology}, 106(5):1773--1794, 2018.

\bibitem{Kalyuzhny2015}
Michael Kalyuzhny, Ronen Kadmon, and Nadav~M. Shnerb.
\newblock A neutral theory with environmental stochasticity explains static and
  dynamic properties of ecological communities.
\newblock {\em Ecology Letters}, 18(6):572--580, 2015.

\bibitem{DANINO2016155}
Matan Danino, Nadav~M. Shnerb, Sandro Azaele, William~E. Kunin, and David~A.
  Kessler.
\newblock The effect of environmental stochasticity on species richness in
  neutral communities.
\newblock {\em Journal of Theoretical Biology}, 409:155--164, 2016.

\bibitem{Hidalgo2017}
Jorge Hidalgo, Samir Suweis, and Amos Maritan.
\newblock {Species coexistence in a neutral dynamics with environmental noise}.
\newblock {\em Journal of Theoretical Biology}, 413:1--10, 2017.

\bibitem{Ashcroft2014}
Peter Ashcroft, Philipp~M Altrock, and Tobias Galla.
\newblock {Fixation in finite populations evolving in fluctuating
  environments.}
\newblock {\em Journal of the Royal Society, Interface}, 11(100):20140663,
  2014.

\bibitem{Cvijovic2015}
Ivana Cvijovic, Benjamin~H Good, Elizabeth~R Jerison, and Michael~M Desai.
\newblock {Fate of a mutation in a fluctuating environment.}
\newblock {\em Proceedings of the National Academy of Sciences},
  112(36):E5021--5028, aug 2015.

\bibitem{melbinger2015impact}
Anna Melbinger and Massimo Vergassola.
\newblock The impact of environmental fluctuations on evolutionary fitness
  functions.
\newblock {\em Scientific reports}, 5(1):1--11, 2015.

\bibitem{Seger1987}
Jon Seger and H.~Jane Brockmann.
\newblock {What is bet-hedging?}
\newblock {\em Oxford Surveys in Evolutionary Biology}, 4:182--211, 1987.

\bibitem{philippi1989hedging}
Tom Philippi and Jon Seger.
\newblock Hedging one's evolutionary bets, revisited.
\newblock {\em Trends in ecology \& evolution}, 4(2):41--44, 1989.

\bibitem{cohen1966optimizing}
Dan Cohen.
\newblock Optimizing reproduction in a randomly varying environment.
\newblock {\em Journal of theoretical biology}, 12(1):119--129, 1966.

\bibitem{Venable2007}
D.~Lawrence Venable.
\newblock {Bet hedging in a guild of desert annuals}.
\newblock {\em Ecology}, 88(5):1086--1090, 2007.

\bibitem{Rajon2014}
Etienne Rajon, Emmanuel Desouhant, Mathieu Chevalier, Fran{\c{c}}ois Debias,
  and Frederic Menu.
\newblock {The evolution of bet hedging in response to local ecological
  conditions.}
\newblock {\em The American naturalist}, 184(1):E1--E15, 2014.

\bibitem{Kussell2005}
Edo Kussell, Roy Kishony, Nathalie~Q Balaban, and Stanislas Leibler.
\newblock {Bacterial persistence: a model of survival in changing
  environments.}
\newblock {\em Genetics}, 169(4):1807--14, 2005.

\bibitem{King2007}
Oliver~D King and Joanna Masel.
\newblock {The evolution of bet-hedging adaptations to rare scenarios.}
\newblock {\em Theoretical population biology}, 72(4):560--75, dec 2007.

\bibitem{libby2019shortsighted}
Eric Libby and William~C Ratcliff.
\newblock Shortsighted evolution constrains the efficacy of long-term bet
  hedging.
\newblock {\em The American Naturalist}, 193(3):409--423, 2019.

\bibitem{Adler2008}
Peter~B. Adler and John~M. Drake.
\newblock Environmental variation, stochastic extinction, and competitive
  coexistence.
\newblock {\em The American Naturalist}, 172(5):E186--E195, 2008.

\bibitem{Pande2020}
Jayant Pande, Tak Fung, Ryan Chisholm, and Nadav~M Shnerb.
\newblock {Mean growth rate when rare is not a reliable metric for persistence
  of species}.
\newblock {\em Ecology Letters}, 23(2):274--282, 2020.

\bibitem{kimura1962probability}
Motoo Kimura.
\newblock On the probability of fixation of mutant genes in a population.
\newblock {\em Genetics}, 47(6):713, 1962.

\bibitem{Laureti2009}
Paolo Laureti, Matus Medo, and Yi~Cheng Zhang.
\newblock {Analysis of Kelly-optimal portfolios}.
\newblock {\em Quantitative Finance}, 10(7):689--697, oct 2010.

\bibitem{Dean2020}
Antony~M Dean and Nadav~M Shnerb.
\newblock {Stochasticity-induced stabilization in ecology and evolution: a new
  synthesis}.
\newblock {\em Ecology}, 101(9):e03098, 2020.

\end{thebibliography}



\end{document}